\documentclass[pra,twocolumn,superscriptaddress]{revtex4}
\usepackage{epsfig}
\usepackage{times}
\usepackage{amsmath}
\usepackage{amsfonts}
\usepackage{amssymb}
\usepackage[usenames]{color}
\usepackage{graphicx}
\newcommand{\beq}{\begin{equation}}
\newcommand{\eeq}{\end{equation}}
\newcommand{\beqa}{\begin{eqnarray}}
\newcommand{\eeqa}{\end{eqnarray}}

\renewcommand{\vec}[1]{\ensuremath{\mathbf{#1}}}

\newcommand{\NTT}{NTT Basic Research Laboratories, NTT Corporation, 3-1
Morinosato-Wakamiya, Atsugi, Kanagawa, 243-0198, Japan.}

\newcommand{\OSAKA}{Graduate School of Engineering Science, University
of Osaka, 1-3 Machikane-yama,
Toyonaka, Osaka 560-8531, Japan.}

\newcommand{\NICT}{National Institute of Information and Communications
Technology, 4-2-1,
Nukuikitamachi, Koganei-city, Tokyo 184-8795 Japan}
\begin{document}
\title{Optically detected magnetic resonance of high-density ensemble of NV$^-$ centers
in diamond.}
 %%%%%%%%%%%%%%%%%%%%%%%%%%%%%%%%%%%%%%%%%%%%%%%%%%%%%%
 % \author{Munro-san?}                                %
 % \affiliation{                                      %
 % NTT Basic Research Laboratories, NTT Corporation,  %
 % Kanagawa, 243-0198, Japan                          %
 % }                                                  %
 %%%%%%%%%%%%%%%%%%%%%%%%%%%%%%%%%%%%%%%%%%%%%%%%%%%%%%

 \author{Yuichiro Matsuzaki}\email{matsuzaki.yuichiro@lab.ntt.co.jp} \affiliation{\NTT}
 \author{Hiroki Morishita}\affiliation{\OSAKA}
 \author{Takaaki Shimooka}\affiliation{\OSAKA}
 \author{Toshiyuki Tashima}\affiliation{\OSAKA}
%\author{Xiaobo Zhu}\affiliation{\china}
 \author{Kosuke Kakuyanagi}\affiliation{\NTT}
 %\author{\newline Kae Nemoto}\affiliation{\NII}
 \author{Kouichi Semba}\affiliation{\NICT}
 \author{W. J. Munro}\affiliation{\NTT}
 \author{Hiroshi Yamaguchi}\affiliation{\NTT}
  \author{Norikazu Mizuochi}\affiliation{\OSAKA}
 \author{Shiro Saito}\affiliation{\NTT}
\begin{abstract}
Optically detected magnetic resonance (ODMR) is a way to characterize
 the NV$^-$ centers. Recently, a remarkably sharp dip was observed
 in the ODMR with a high-density ensemble of NV centers, and this was
 reproduced
 by a theoretical model in [Zhu {\it{et
 al}}., Nature Communications {\bf{5}}, 3424 (2014)], showing that
 the dip is a consequence of the spin-1 properties of the NV$^-$
 centers. 
 Here, 
 we present much more details of analysis
 to show how this model can be applied to investigate the properties
 of the NV$^-$ centers.
%  Here, we perform ODMR on a high-density NV centers ensemble with
%  and without magnetic field. We construct a model to reproduce the
%  experimental results.
 \textcolor{black}{By using our model, we have reproduced the ODMR with and without applied magnetic
 fields. Also, we theoretically investigate how the ODMR is affected by the typical
 parameters of the ensemble NV$^-$ centers such as strain distributions,
 inhomogeneous magnetic fields, and homogeneous broadening width. Our model could provide a way to estimate these parameters
 from the ODMR, which would be crucial to realize diamond-based quantum
 information processing. 
 }
% We have found that our approach can be applied to determine the typical
% parameters of the ensemble of NV centers; strain distributions,
 %inhomogeneous magnetic fields, and homogeneous broadening width.
 %Such a parameter estimation is essential for the use of NV$^-$ centers
 %to realize diamond-based quantum information processing. 
\end{abstract}
\maketitle

\section{Introduction}

\textcolor{black}{
 A nitrogen-vacancy (NV$^-$) center in diamond \cite{davies1976optical,gruber1997scanning,Go01a} is a promising candidate 
 to realize quantum information processing \cite{dutt2007quantum,
wrachtrup2001quantum,jelezko2002single,NMRHWYJGJW01a,robledo2011high,shimo2015control} and network \cite{CTSL01a}.
\textcolor{black}{An NV$^-$ center is known to have a long coherence time such as a
 second \cite{mizuochi2009coherence,balasubramanian2009ultralong,bar2013solid}.}
 The operations such as 
 qubit gates and measurements, which are basic tools for quantum 
 applications, have been demonstrated with a single NV center  \cite{JGPGW01a}. 
 Also, the entanglement generation between distant nodes, which plays an essential role of 
 quantum repeater, has been demonstrated by using photons as flying 
 qubits emitted through distant two single NV centers
 \cite{bernien2013heralded}.
 \textcolor{black}{An NV$^-$ center can be used for a sensitive magnetic field sensor
 \cite{maze2008nanoscale, taylor2008high,
balasubramanian2008nanoscale}.}
 An ensemble 
 of NV$^-$ centers can be also used for demonstrating quantum metrology
   \cite{pham2011magnetic,acosta2009diamonds,steinert2010high}
 and physical phenomena in fundamental physics, such as quantum walk 
  \cite{hardal2013discrete}, and quantum simulation
 \cite{yang2012quantum}. Also, the ensemble of NV$^-$ centers can be used as the hybrid 
 devices between different physical systems, in particular, superconducting 
 systems \cite{imamouglu2009cavity,wesenberg2009quantumetal,schuster2010highetal,kubo2010strongetal,
amsuss2011cavityetal,marcos2010couplingetal,julsgaard2013quantumetal,diniz2011stronglyetal,putz2014protecting,zhu2011coherent,zhudark2014,
kubo2011hybridetal,saito2013towards}. Due
 to the effect of a superradiance,
 the ensemble of 
 NV$^-$ centers has a much stronger magnetic coupling with other systems 
 than a single NV$^-$ center.
}

\textcolor{black}{
 An NV$^-$ center consists of a nitrogen atom and a vacancy in the 
 adjacent site \cite{davies1976optical}, and this is a spin-1 system with three states of 
 $|0\rangle $, $|-1\rangle $, and $|1\rangle $. With a strong external magnetic 
 field, the two exited states $|1\rangle $ and $|-1\rangle $ of the 
 NV$^-$ center is energetically separated far from each other. In this 
 case, the NV$^-$ center can be considered as a spin $\frac{1}{2}$ system by 
 using a frequency selectivity where $|0\rangle $ and $|1\rangle $
 ($|-1\rangle $) constitute a qubit. On the other hand, with zero or weak 
 applied magnetic field, the NV$^-$ center reveals spin-1 properties 
  \cite{alegre2007polarization,fang2013high,dolde2011electric}. Optically detected magnetic resonance (ODMR) is the general 
 technique to investigate the properties of the NV$^-$ centers \cite{gruber1997scanning}. After 
 applying a microwave pulse, the NV$^-$ centers are measured by an 
 optical detection. Resonance observed with specific microwave 
 frequencies let us know an energy structure of a ground-state manifold 
 of the NV$^-$ centers. Also, we can estimate coherence properties of the 
 NV$^-$ center from the width of the peaks.
}

Recently, a remarkably sharp dip has been observed around
 2870 MHz in the ODMR with zero applied magnetic fields
 \cite{kubo2010strongetal,simanovskaia2013sidebands,zhudark2014}.
 \textcolor{black}{Although the ODMR results are usually fit by a sum of Lorentzians,}
  the ODMR results observed in \cite{kubo2010strongetal,simanovskaia2013sidebands,zhudark2014} cannot be well
 reproduced by such a fitting
 %a sum of Lorentzians
 \cite{kubo2010strongetal}, and
 no theoretical model can explain the dip until
  a new approach is suggested in \cite{zhudark2014}.
  The model described in \cite{zhudark2014} contains spin-1 properties
 of the NV$^-$ centers while most of the previous models assume the
 NV$^-$ center to be a spin-half system \textcolor{black}{or use just a
 sum of Lorentzians to include the effect of the spin-1 properties in a
 phenomenological way \cite{kubo2010strongetal}}. By including the strain
 distributions, randomized magnetic fields, and homogeneous width of the
 NV$^-$ centers, the sharp dip in the ODMR has been reproduced in
 \cite{zhudark2014}.
 \textcolor{black}{This model provides us with an efficient tool to
 characterize the high-density ensemble of NV$^-$ centers, which would
 be crucial to realize diamond-based quantum information processing.}
 Moreover, this dip is shown to be
 the cause
 of a long-lived collective dark state observed in a spectroscopy of
 superconductor diamond hybrid system, and so this dip could be useful
 if we will use the collective dark state for a long-lived quantum memory of a
 superconducting qubit \cite{zhudark2014}.

 In this paper, we present the details about how the model suggested in
 \cite{zhudark2014} can be applied to
 investigate the properties of an ensemble of NV$^-$ centers.
 \textcolor{black}{
  An ensemble of NV$^-$ centers is affected by inhomogeneous magnetic
 fields, inhomogeneous strain distributions, and homogeneous broadening.
 By taking into account of these as parameters in our model, we have reproduced the ODMR with and without applied
 magnetic field.
Moreover, from a numerical simulation, we have investigated how these parameters
affect the sharp dip around 2870 MHZ and the width of the each peak in
 the ODMR. We have found that homogeneous broadening is relevant to change
 the dip in the ODMR. Also, we have confirmed that the width of the peaks in the
 ODMR is insensitive against the strain variations if an external
 magnetic field is applied. Moreover, we have shown how our model could
 be used to estimate these parameters of the NV$^-$ centers from the ODMR.}
 % An ensemble of NV$^-$ centers is affected by inhomogeneous magnetic
%  fields, inhomogeneous strain distributions, and homogeneous broadening.
% In the results of the ODMR, the observed peaks
%  contain the information of the total width that is a composite effect of
%  three noise mentioned above, and it is not straightforward to separate
%  these three effects for the
%  estimation about how individual noise contributes to the width.
%  Here, by using our model to reproduce the ODMR, we construct a way how we estimate the
%  inhomogeneous magnetic-field
%  width, inhomogeneous strain width, and homogeneous width of the
%  NV$^-$ center, respectively. This technique provides an efficient way to characterize the ensemble
%  of NV$^-$ centers, and is particular useful when we try to use the
%  spin-1 properties of the NV$^-$ center for quantum information processing.

The rest of this paper is organized as follows. In section 2, we explain
the experimental setup.
In section 3,
we introduce the theoretical model introduced in \cite{zhudark2014}.
In section 4, we show the ODMR results and explain how these
experimental results can be reproduced by our theoretical model. Finally, section 5 contains a summary
of our results.

\section{Experimental setup}
We begin by describing how we generate the NV$^-$ centers in diamond.
To create the NV$^-$ center ensemble, we performed ion implantation of
$^{12}$C$^{2+}$
%at
%700 keV into type Ib (001) diamond
%with a nominal nitrogen concentration of 100 ppm
 and we annealed the sample in
high vacuum \cite{zhu2011coherent}.
%The $NV^-$ ensemble is generated by an ion implantation and annealing in
%vacuum .
The density of the NV$^-$ centers is approximately
$5\times 10^{17}$ cm$^{-3}$, and we have the NV$^-$ centers over the
depth of 1$\mu$m from the surface of the diamond.

\begin{figure}[h!]
\includegraphics[scale=0.17]{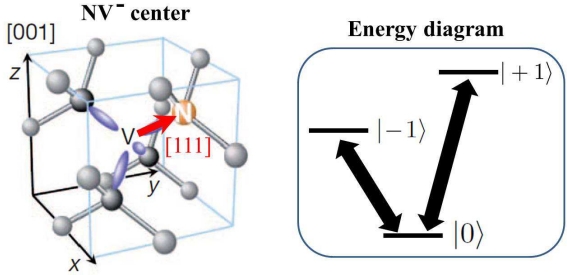}%{figure-dark-threed}%
\caption{
NV$^-$ center
 consists of a nitrogen atom (N)
 and a vacancy (V) in the adjacent site. Since NV$^-$ center is a spin-1
 system, we have three states of $|0\rangle $, $|1\rangle $, and
 $|-1\rangle$.
 We can characterize the NV$^-$ center by
 an optically detected magnetic
 resonance (ODMR) spectrum, and we perform the ODMR with an applied
 magnetic field of $B=0,1,2$ mT, along the [111] direction.
 }
\label{config}%
\end{figure}
The ODMR was performed on the diamond sample
by a confocal microscope with a magnetic resonance system
at room temperature \cite{mizuochi2009coherence}.
\textcolor{black}{We manipulate pulsed optical laser (532nm) and microwave independently.}
% We irradiate the
% optical laser
% (532 nm) and microwave separately by using a pulsed technique.
The magnetic field of $0$, $1$, or $2$ mT was applied along the [111]
axis.
%In the diamond crystal structure,
With zero or weak applied magnetic field,
a quantization
axis of the NV$^-$ center is determined by the direction from the vacancy to
the nitrogen, which we call an NV$^-$ axis. This axis is along one
of four possible crystallographic axes.
%We call the vacancy-nitrogen direction an ``NV$^-$ axis''. 
%such as [1,1,1],
%[$\overline{1}$,1,1], [1,$\overline{1}$,1], and [1,1,$\overline{1}$].
The NV$^-$ centers usually
occupy these four directions equally.
The applied magnetic field
along [111] is aligned with one of these four axes as shown in
Fig. \ref{config}. In this case, the Zeeman
splitting of the NV$^-$
centers having the NV$^-$ axis of [111] is larger than that of the NV$^-$ centers having
the other three NV$^-$ axes.
%On the other hand, the NV$^-$
%centers with the other three axes are affected by the same amount of the Zeeman splitting.

\section{Model}
We describe the model to simulate the ODMR of the NV$^-$ center
ensemble,
which was introduced in \cite{zhudark2014}.
\textcolor{black}{The NV$^-$ axis provides us with the $z$ axis.
Microwave pulses are applied on the NV$^-$ centers, and the microwave
pulses orthogonal to the z axis induce the excitation of the NV$^-$ centers.
We define the $x$ axis as such a \textcolor{black}{orthogonal direction} of the
applied microwave at each NV$^-$ center.}
The Hamiltonian of the NV$^-$ centers is as follows.
%The Hamiltonian is as follows.
\textcolor{black}{
\begin{eqnarray}
 H&=&\hbar
  \sum_{k=1}^{N}\Big{\{}D_k\hat{S}^2_{z,k}+g_e\mu _BB^{(k)}_z \hat{S}_{z,k}+E_{1}^{(k)}(\hat{S}_{x,k}^2-\hat{S}_{y,k}^2)
  \nonumber \\
  &+&E_{2}^{(k)}(\hat{S}_{x,k}\hat{S}_{y,k}+\hat{S}_{y,k}\hat{S}_{x,k})
 +\lambda \cos (\omega t) \hat{S}^{(k)}_{x}\nonumber \\
&+&A_{\|}\hat{S}_{z,k}\hat{I}_{z,k}+\frac{A_{\perp}}{2}(\hat{S}_{+,k}\hat{I}_{-,k}+\hat{S}_{-,k}\hat{I}_{+,k})\nonumber \\
 &+& P(\hat{I}_{z,k}^2-\frac{1}{3}\hat{I}^2)
 -g_{\text{n}}\mu _{N}B^{(k)}_{z}\hat{I}_{z,k}
  \Big{\}}\nonumber
\end{eqnarray}
where $\hat{S}_k$ ($\hat{I}_k$) denotes a spin-1 operator of
$k$th electron (nuclear) spin, $D_k$ denotes a zero-field splitting, $E^{(k)}_1$ ($E^{(k)}_2$) denotes a strain
along x(y) direction, $g_e\mu _BB^{(k)}_{z}\cdot \vec{S}_k$
($-g_{\text{n}}\mu _{N}B^{(k)}_z I_{z,k}$) denotes a Zeeman term of
the $k$th electron (nuclear) spin,
$\lambda $ denotes a microwave amplitude, $\omega $ denotes a
microwave frequency, $P$ denotes the quadrupole splitting, and $A_{\|}$ ($A_{\perp}$) denotes a parallel
(orthogonal) hyperfine coupling. \textcolor{black}{For simplicity, we assume a homogeneous
microwave amplitude $\lambda _k\simeq \lambda$. (In the appendix, we
relax this constraint.)}
It is worth mentioning that
  the x and y component of the magnetic field is
  insignificant to change
  quantized axis
  %of the NV$^-$ center for a small magnetic field $D\gg
  %g_e\mu_B |{\bf {B}}_{\text{NV}}|$,
  and so we consider only the effect
  of z axis of the
  magnetic field.
  Since the energy of the nuclear spin is detuned from the energy of the
  electron spin, the flip-flop term
  $\frac{A_{\perp}}{2}(\hat{S}_{+,k}\hat{I}_{-,k}+\hat{S}_{-,k}\hat{I}_{+,k})$
  is negligible and the parallel term
  $\hat{A}_{\|}\hat{S}_{z,k}\hat{I}_{z,k}$ is dominant. In this case, the effect of the nuclear spin is
  considered as randomized magnetic fields on the electron spin
  \cite{kubo2011hybridetal, saito2013towards}, and we use this
  approximation throughout the paper.}
  
  In a rotating frame 
  defined by
 $U=e^{-i\omega
 \hat{S}_z^2t/\hbar}$, we can perform the rotating wave approximation, and we
 obtain the simplified Hamiltonian.
 \begin{eqnarray}
 H\simeq \hbar
  \sum_{k=1}^{N}\Big{\{}(D_k-\omega )\hat{S}^2_{z,k}+E_{1}^{(k)}(\hat{S}_{x,k}^2-\hat{S}_{y,k}^2)
  \ \ \ \ \ \ \ \ \ \ \ 
  \nonumber \\
  +E_{2}^{(k)}(\hat{S}_{x,k}\hat{S}_{y,k}+\hat{S}_{y,k}\hat{S}_{x,k})
 +g_e\mu _BB^{(k)}_z\hat{S}_z+\frac{\lambda}{2}\hat{S}^{(k)}_{x}
  \Big{\}}\nonumber
\end{eqnarray}
If the number of excitations in the spin ensemble is much smaller than the
number of spins, we can consider the spin ensemble as a number of harmonic
oscillators. In this case, we can replace the spin ladder operators as 
creational operators of the harmonic oscillators such as
$\hat{b}^{\dagger }_k\simeq |B\rangle _k\langle 0|$, $\hat{d}^{\dagger
}_k\simeq |D\rangle _k\langle 0|$ where $|B\rangle _k=\frac{1}{\sqrt{2}}(|1\rangle_k
+|-1\rangle _k)$, $|D\rangle _k=\frac{1}{\sqrt{2}}(|1\rangle_k
-|-1\rangle _k)$.
By using this approximation, we can simplify the Hamiltonian as follows \cite{zhudark2014}.
\begin{eqnarray}
 H&\simeq &\hbar
  \sum_{k=1}^{N}
  \Big{\{}(\omega ^{(k)}_{b}-\omega )\hat{b}^{\dagger }_k\hat{b}_k
  +(\omega ^{(k)}_{d}-\omega )\hat{d}^{\dagger }_k\hat{d}_k\nonumber \\
  &+&J_k(\hat{b}^{\dagger
  }_k\hat{d}_k+\hat{b}_k\hat{d}^{\dagger }_k
  )
 +iJ'_k(\hat{b}^{\dagger
  }_k\hat{d}_k-\hat{b}_k\hat{d}^{\dagger }_k
  )+\frac{\lambda }{2}(\hat{b}_k+\hat{b}^{\dagger }_k)
  \Big{\}}\nonumber
\end{eqnarray}
where $\omega ^{(k)}_b=D_k -E^{(k)}_{1}$, $\omega ^{(k)}_d=D_k +E^{(k)}_{1}$,
$J_k=g\mu _BB^{(k)}_z$, and $J'_k=E^{(k)}_{2}$.

The inhomogeneous broadening can be included in this model as following.
We use Lorentzian distributions to include an inhomogeneous
effect of \textcolor{black}{$E^{(k)}_{1}$, and $E^{(k)}_{2}$}
$(k=1,2,\cdots ,N)$. It is worth mentioning that the Lorentzian
distributions have
 been typically used to
describe the inhomogeneous broadening of the NV$^-$ centers
\cite{zhu2011coherent,zhudark2014,kubo2011hybridetal,kubo2012electronetal}.
For an inhomogeneous magnetic field $B^{(k)}_z$, we need to consider the
following two effect. 
 First,
 %due to the electron spin-half bath in the environment such as
 since there is an electron spin-half bath in the environment due to the
 \textcolor{black}{substitutional N  (P1)} centers,
 NV$^-$ centers are affected by randomized magnetic fields. Second, a hyperfine
coupling of the nitrogen nuclear spin
splits the energy of the NV$^-$ center into three levels. So
we use a random distribution of the magnetic fields with
the form of
the mixture of three Lorentzian functions. Here, each peak of the Lorentzian is separated with $2\pi \times
2.3 $ MHz that corresponds to the hyperfine interaction with \textcolor{black}{$^{14}\mathrm{N}$}
 nuclear spin
 \cite{kubo2011hybridetal, saito2013towards}.
 It is worth mentioning that, since the
frequency shift of $D_k$ is almost two-orders of magnitude smaller than
 that of $E^{(1)}_k$ and $E^{(2)}_k$ \cite{dolde2011electric}, we consider the effect
 of inhomogeneity of $D_k$ as this order in this paper.

We can describe the dynamics of the NV$^-$ centers by using the Heisenberg equation as
follows.
\begin{eqnarray}
 \frac{d\hat{b}_k}{dt}&=&-i(\omega ^{(k)}_b-i\Gamma _b)\hat{b}_k-iJ_k
  \hat{d}_k+J'_k\hat{d}_k-i\lambda \nonumber \\
  \frac{d\hat{d}_k}{dt}&=&-i(\omega ^{(k)}_d-i\Gamma _d)\hat{d}_k-iJ_k
  \hat{b}_k-J'_k\hat{b}_k 
\end{eqnarray}
where $\Gamma _b(=\Gamma _d)$ denotes the homogeneous width of the NV
center. We assume that the initial state is a vacuum state. Since we consider a steady state after a long time, we can set the time
derivative as zero. In this condition, we obtain
\begin{eqnarray}
\langle \hat{b}^{\dagger }_{k,t=\infty } \hat{b}_{k,t=\infty
 }\rangle\ \ \ \ \ \ \ \ \ \ \ \ \ \ \ \ \ \ \ \ \ \ \ \ \ \
 \ \ \ \ \ \ \ \ \ \ \ \ \ \ \ \ \ \ \ \ \ \ \ \ \ \ \ \ \ \ \ \ \ \ \ \ \ \ \ 
 \nonumber \\
 = |\frac{\lambda(\omega -\omega ^{(k)}_d+i\Gamma _d)}{(\omega -\omega ^{(k)}_b+i\Gamma
  _b)(\omega -\omega ^{(k)}_d+i\Gamma _d) -(|J_k|^2
  +|J'_k|^2)}|^2\label{brightex} \\
  \langle \hat{d}^{\dagger }_{k,t=\infty }\hat{d}_{k,t=\infty }\rangle
   \ \ \ \ \ \ \ \ \ \ \ \ \ \ \ \ \ \ \ \ \ \ \ \ \ \ \ \ \ \ \ \ \ \ \
   \ \ \ \
   \ \ \ \ \ \ \ \ \ \ \ \ \ \ \ \ \ \ \ \ \ \ \ \ \ \ 
   \nonumber
    \\
   =| \frac{\lambda(J_k-iJ'_k)}{(\omega -\omega ^{(k)}_b+i\Gamma
  _b)(\omega -\omega ^{(k)}_d+i\Gamma _d) -(|J_k|^2 +|J'_k|^2)}|^2 \label{darkex}
\end{eqnarray}
% \textcolor{red}{The average probability of the NV$^-$ center to be in the
% ground states ($|0\rangle $) can be calculated as}
% \begin{eqnarray}
%  P_g= \frac{1}{N}(\sum_{k=1}^{N} 1-\langle \hat{b}^{\dagger }_{k,t=\infty
%   }\hat{b}_{k,t=\infty }\rangle
%   -\langle \hat{d}^{\dagger }_{k,t=\infty }\hat{d}_{k,t=\infty }\rangle
%   ) \label{probabilityg}.
% \end{eqnarray}
\textcolor{black}{The average probability of the NV$^-$ center to be in the
energy eigenstates other than $|0\rangle $ can be calculated as}
\begin{eqnarray}
 P_e= \frac{1}{N}(\sum_{k=1}^{N} \langle \hat{b}^{\dagger }_{k,t=\infty
  }\hat{b}_{k,t=\infty }\rangle
  +\langle \hat{d}^{\dagger }_{k,t=\infty }\hat{d}_{k,t=\infty }\rangle
  ) \label{probabilitye}.
\end{eqnarray}

In the actual experiment, if we excite
the NV$^-$
centers by the microwave pulses,
the intensity of the photons emitted from the
NV$^-$ centers will be changed from \textcolor{black}{the baseline emission rate $I_0$. This change
is linear with $P_e$. So, to fit the experiment with our
model, we use a function of $(I_0-aP_e)/I_0$
% \begin{eqnarray}
%  \frac{I_0-aP_e}{I_0}
% \end{eqnarray}
% \begin{eqnarray}
% a \Big{(} \frac{1}{N}(\sum_{k=1}^{N} 1-\langle \hat{b}^{\dagger }_{k,t=\infty
%   }\hat{b}_{k,t=\infty }\rangle
%   +\langle \hat{d}^{\dagger }_{k,t=\infty }\hat{d}_{k,t=\infty }\rangle )\Big{)}+b
% \end{eqnarray}
where $a$ denotes a fitting parameter
, and this corresponds to the ODMR signals.}

\section{Main results}
\subsection{Reproducing the experimental results}
  \begin{figure}[h!]
\includegraphics[scale=0.19]{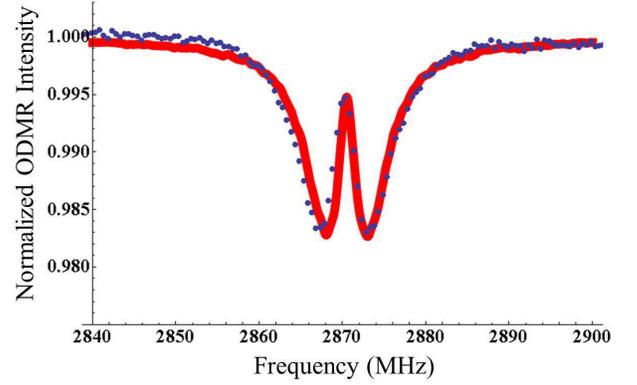}%{figure-dark-threed}%
\caption{
ODMR with zero applied magnetic fields.
   %$N=12000$,
   $\delta (g\mu _BB _z)/2\pi =1.96$ MHz (HWHM), $\delta E_1/2\pi =\delta E_2/2\pi=0.73$ 
   MHz (HWHM),
   $\delta D_k=0.01$ MHz (HWHM),
   $\lambda /2\pi=2$ MHz, $\Gamma _b/2\pi=\Gamma _d/2\pi= 0.3$ MHz. Also,
   we assume a Nitrogen hyperfine coupling of $2\pi \times 2.3$ MHz. The
   red line denotes a numerical simulation and blue dots denote the
   experimental results.
%See 8allexpodmr.nb. C: Users yuichiro Documents 0ntt15-4-doc odmr-paper simulation-odmr for-figure
 }
\label{zero}%
\end{figure}
  \begin{figure}[h!]
\includegraphics[scale=0.19]{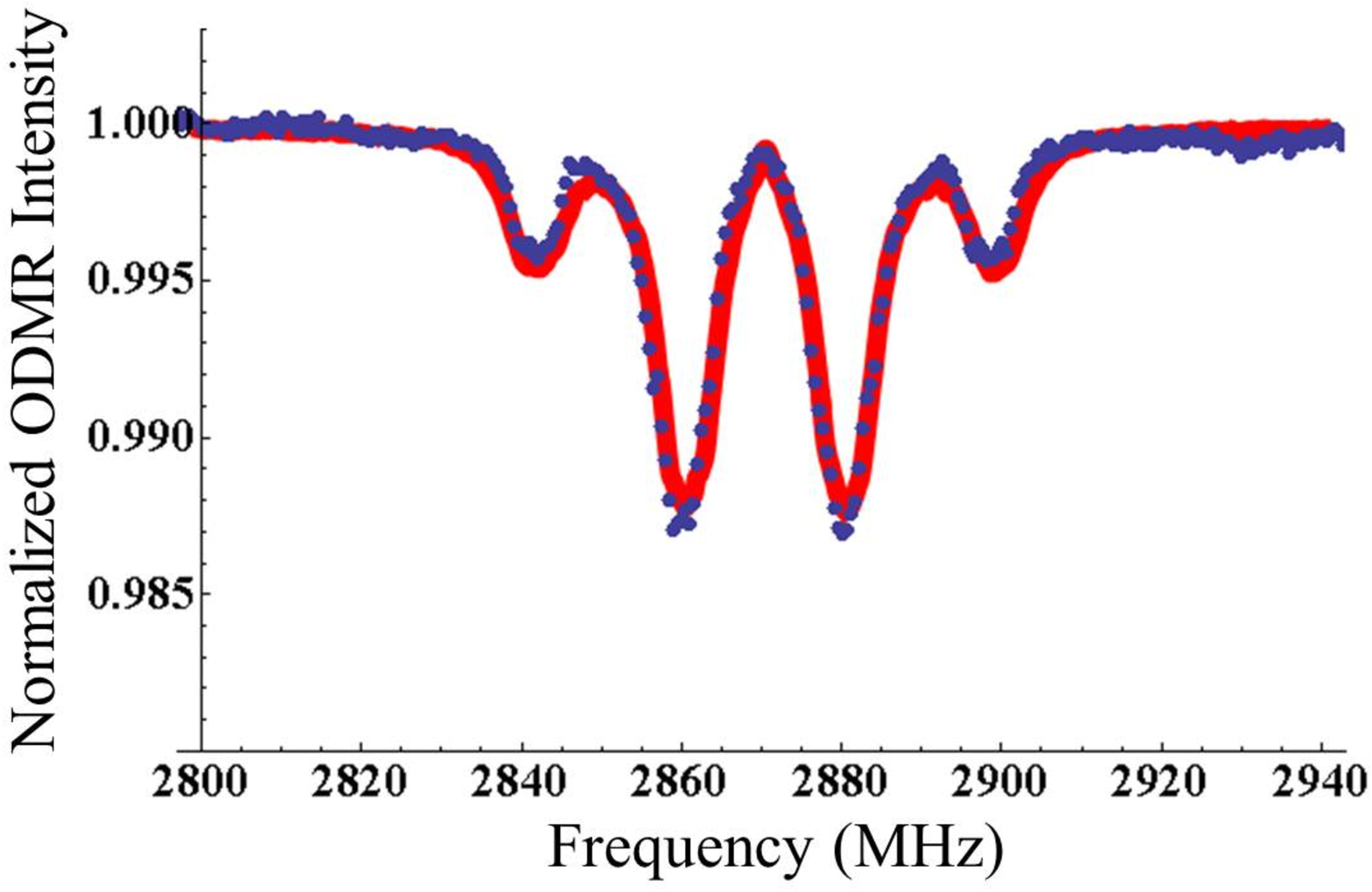}%{figure-dark-threed}%
\caption{
ODMR with an applied magnetic field of 1 mT. We use the same parameters
   as the Fig. \ref{zero}. The
   red line denotes a numerical simulation and blue dots denote the
   experimental results.
%See 8allexpodmr.nb. C: Users yuichiro Documents 0ntt15-4-doc odmr-paper simulation-odmr for-figure
 }
\label{onemt}%
\end{figure}
  \begin{figure}[h!]
\includegraphics[scale=0.19]{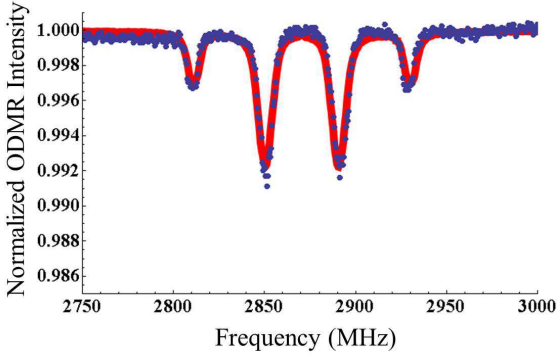}%{figure-dark-threed}%
\caption{
ODMR with an applied magnetic field of 2 mT. We use the same parameters
   as the Fig. \ref{zero}. The
   red line denotes a numerical simulation and blue dots denote the
   experimental results.
%See 8allexpodmr.nb. C: Users yuichiro Documents 0ntt15-4-doc odmr-paper simulation-odmr for-figure
 }
\label{twomt}%
\end{figure}
By using the model described above, we have reproduced the ODMR signals when we apply $B=0,1,2$ mT, as shown
in Figs \ref{zero}, \ref{onemt}, and \ref{twomt}. A sharp dip is
observed around $2870$ MHz for the case of $B=0$ mT, and our
simulation can reproduce this.
% We explain the reason why the sharp dip is observed in the ODMR with
% zero applied magnetic field.
 Two peaks are observed in the ODMR with zero applied magnetic field as
 shown in Fig \ref{zero}, which corresponds to the transition between the state \textcolor{black}{$|0\rangle$ and one
 of the other energy eigenstates}. If we consider a single NV$^-$ center,
 the frequency difference between the two exited states is
 $\delta \omega =2\sqrt{(g\mu _BB_z)^2 +(E_1)^2+(E_2)^2}$.
Since we consider an ensemble of the NV$^-$ center, this frequency
difference
variates depending on the position of the NV$^-$ center.
For simplicity, we use a dimensionless variable for $B_z$, $E_1$, and
$E_2$, defined as $\tilde{B}_z=g\mu _BB_z/\gamma $, $\tilde{E}_1=E_1/\gamma $, and
$\tilde{E}_2=E_2/\gamma $ where $\gamma $ denotes a damping rate with an unit of the frequency.
To calculate a probability that the \textcolor{black}{two energy
eigenstates such as
$|1\rangle $ and $|-1\rangle $} are degenerate
($\delta \omega =0$), we define probability density functions of $\tilde{B}_z$,
$\tilde{E}_1$, and $\tilde{E}_2$ as $P_a(\tilde{B}_z)$,
$P_b(\tilde{E}_1)$, and $P_c(\tilde{E}_2)$,
respectively.
The joint probability is calculated as
\begin{eqnarray}
&& P(\tilde{B}_z,\tilde{E_1},\tilde{E_2})\Delta \tilde{B}_z\Delta
 \tilde{E}_1\Delta \tilde{E}_2\nonumber \\
  &=&P_a(\tilde{B}_z=0)\Delta \tilde{B}_z\cdot P_b(\tilde{E}_1=0)\Delta
   \tilde{E}_1\cdot
   P_c(\tilde{E}_2=0)\Delta \tilde{E}_2.\nonumber
\end{eqnarray}
where $\Delta \tilde{B}_z$, $\Delta E_1$, and $\Delta E_2$ denote a
finite range of each variable. We assume
$P(\tilde{B}_z,\tilde{E_1},\tilde{E_2})=P_a(\tilde{B}_z)P_b(\tilde{E}_1)P_c(\tilde{E}_2)$
because these are independent.
By using spherical coordinates where $\tilde{B}_z=r \sin \theta \cos
\phi $, $\tilde{E}_1=r \sin \theta \sin \phi$, $\tilde{E_2}=r \cos
\theta $ with $r=\sqrt{|\tilde{B}_z|^2+|\tilde{E}_1|^2+|\tilde{E}_2|^2}$, we rewrite this as
\begin{eqnarray}
&& P(\tilde{B}_z,\tilde{E_1},\tilde{E_2})\Delta \tilde{B}_Z\Delta
 \tilde{E}_1\Delta \tilde{E}_2\nonumber \\
  &=&P_a(\tilde{B}_z=0) P_b(\tilde{E}_1=0) P_c(\tilde{E}_2=0)r^2 \sin
   \theta \Delta r \Delta \theta \Delta \phi. \nonumber
\end{eqnarray}
This shows that, even if we consider a finite range 
$\Delta \tilde{B}_z$, $\Delta E_1$, and $\Delta E_2$, the probability
for the two energy \textcolor{black}{eigenstates} to be exactly degenerate ($r=0$) is zero.
This means that, if homogeneous broadening is negligible, the two peaks to denote the two
energy \textcolor{black}{eigenstates} of each NV$^-$ center should be always separated in the ODMR
so that the ODMR signal at the frequency
of $D/2\pi =2870$ MHz should be the same as the base line.
However, due to the effect of the homogeneous broadening, small signals
deviated from the base line
can be observed at the frequency
of $D/2\pi =2870$ MHz. This is the cause of the sharp dip observed
around the frequency of $2\pi \times 2870$ MHz in
the ODMR.

With an applied magnetic field, four peaks are observed in the ODMR where two of
them are larger than the other two, as shown in Figs
\ref{onemt} and \ref{twomt}.
The two smaller peaks
correspond to the energy  \textcolor{black}{eigenstates} of the NV$^-$
centers with an NV$^-$ axe along [111], which is aligned with the
applied magnetic field. A quarter of the NV$^-$ centers in the ensemble have such an
NV$^-$ axis.
The other larger peaks come from the other NV$^-$ centers where the
applied magnetic field is not aligned with the NV$^-$ axis.
Three-quarters NV$^-$ center have such axes.
In this case,
the Zeeman splitting  of these is smaller than that of the NV$^-$
centers with the [111] axis. It is worth mentioning that a small dip is
observed in the 1mT ODMR around $2\pi \times 2870$ MHz  due to the
mechanism explained above. On the other hand, such a dip is not
clearly observed
in the 2mT ODMR, because the NV$^-$ centers
are considered to be as approximate two-level systems in this regime.

\subsection{The behavior of the ODMR against the change in the parameters}
\begin{figure}[h!]
\includegraphics[scale=0.25]{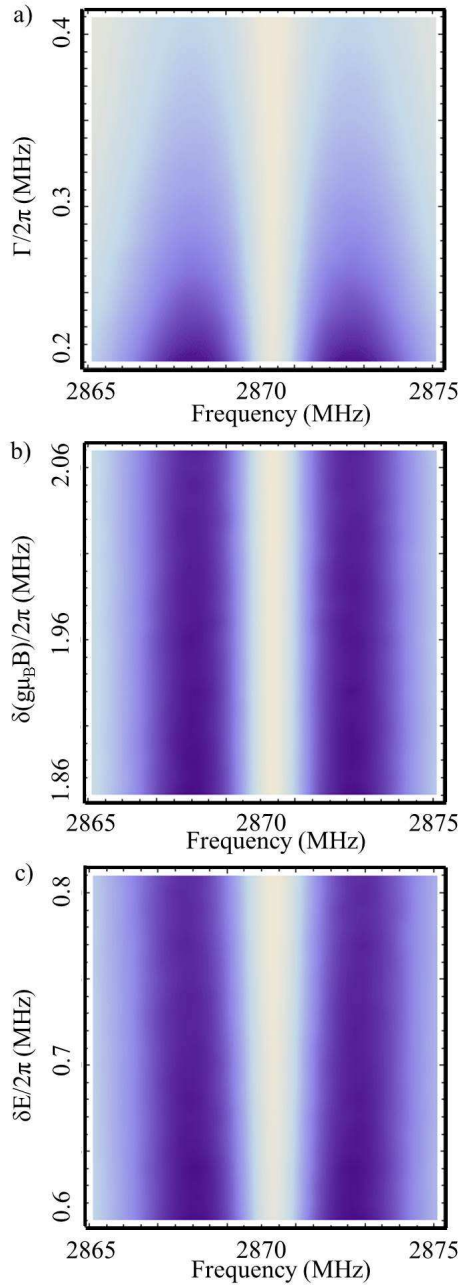}%{figure-dark-threed}%
\caption{
Numerical simulation of ODMR with zero applied magnetic fields.
    Here, $x$ axis denotes the microwave
   frequency, $y$ axis denotes $\gamma /2\pi $ (for the figure a) or $\delta (g\mu
 _BB_z)/2\pi $ (for the figure b)
 or $\delta E/2\pi $ (for the figure c), and $z$ axis
   denotes the ODMR signal intensity.
   Other
    than the inhomogeneous width of the inhomogeneous width, we use the same parameters
    as the Fig. \ref{zero}.
% We use $\delta E_1=\delta E_2= 0.6,0.8,1.0$ MHz  (green, red, blue).
%See 7strainchange.nb C: Users yuichiro Documents 0ntt15-4-doc odmr-paper simulation-odmr for-figure
 }
\label{zeroebg}%
\end{figure}

We perform a numerical simulation with several parameters to
understand the behavior of the sharp dip.
In the Fig \ref{zeroebg} a,  we change the
parameter $\Gamma _{b}$
while we fix the other parameters. Similarly, in the Fig \ref{zeroebg} b
(c),
we change the parameter $\delta B_k$ ($\delta E_k$) while we fix the
other parameters. We have found that the sharp dip is very sensitive
against the change in $\Gamma _{b}$, while the dip is
relatively insensitive against the change in $\delta E_k$ and $\delta
B_k$.

\begin{figure}[h!]
\includegraphics[scale=0.175]{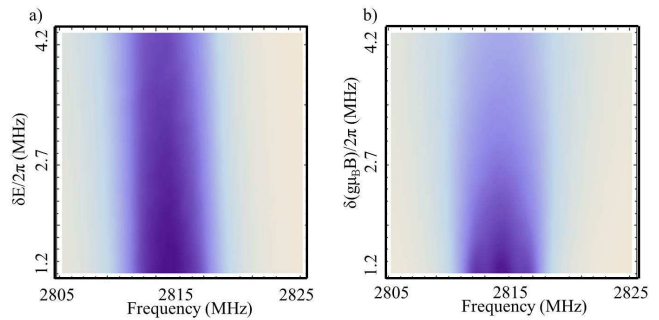}%{figure-dark-threed}%
\caption{
Numerical simulation of ODMR with an applied magnetic field of 2 mT.
     Here, $x$ axis denotes the microwave
   frequency, $y$ axis denotes $\delta E/2\pi $ for the left figure while $y$
 axis denotes $\delta (g\mu _BB)/2\pi $ for the right figure, and $z$ axis
   denotes the ODMR signal intensity.
 Other
   than $\delta E$ or $\delta (g\mu
 _BB_z)/2\pi $, we use the same parameters
   as the Fig. \ref{twomt}. \textcolor{black}{These show that the peak width is much more sensitive against
 the inhomogeneous magnetic fields than inhomogeneous strain.}
 %We use $\delta B_k= 0.0414,0.700,0.0986$ mT  (green, red, blue).
%See b-bchange.nb C: Users yuichiro Documents 0ntt15-4-doc odmr-paper simulation-odmr for-figure
 }
\label{bb}%
\end{figure}

%\subsection{Behavior of the ODMR with applied magnetic field.}
Also, we perform a numerical simulation with several parameters for the
ODMR with an applied magnetic field.
In the Figs \ref{bb}, we have plotted one of the peaks of the
ODMR with an applied magnetic field of 2mT. This peak corresponds to a
transition between $|0\rangle $ and $|-1\rangle $ of the NV$^-$ center
with an axis of [111].
% \begin{figure}[h!]
% \includegraphics[scale=0.2]{2b-bchange.eps}%{figure-dark-threed}%
% \caption{
% Numerical simulation of ODMR with an applied magnetic field of 2 mT. Other
%    than $\delta B_k$, we use the same parameters
%    as the Fig. \ref{zero}. We use $\delta B_k= 0.0414,0.700,0.0986$ mT
%    (green, red, blue).
% See b-bchange.nb C: Users yuichiro Documents 0ntt15-4-doc odmr-paper simulation-odmr for-figure
%  }
% \label{bb}%
% \end{figure}
% \begin{figure}[h!]
% \includegraphics[scale=0.2]{2b-strainchange.eps}%{figure-dark-threed}%
% \caption{
% Numerical simulation of ODMR with an applied magnetic field of 2 mT. Other
%    than $\delta E$, we use the same parameters
%    as the Fig. \ref{zero}. We use $\delta E= 0,0.8,1.6$ MHz
%    (green, red, blue).
% See b-bchange.nb C: Users yuichiro Documents 0ntt15-4-doc odmr-paper simulation-odmr for-figure
%  }
% \label{be}
% \end{figure}
From the numerical simulations, we have found that this ODMR
signals with applied magnetic field is robust against the
strain variations $\delta E_k$, while the peak will be broadened due to the effect of the
randomized magnetic field $\delta B_k$.
The frequency difference between the ground state and
\textcolor{black}{another energy eigenstate}
can be calculated as $\delta \omega '=D_k-
\sqrt{E^{(k)}_{1}+E^{(k)}_{2}+(g_e\mu _BB_z^{(k)})^2}$. If the applied
magnetic field is large, we obtain $\delta \omega '\simeq g_e \mu _B B^{(k)}_z\pm \frac{|E^{(k)}_{1}|^2+|E^{(k)}_{2}|^2}{2g_e \mu
_B B^{(k)}_z}$. This means that the effect of the strain is
insignificant in this regime while the inhomogeneous magnetic field
from the environment can easily change this frequency. These can
explain the simulation results shown in Figs \ref{bb}
\textcolor{black}{where the change of inhomogeneous magnetic fields
affects the width of the peak while the peak width is insensitive
against the inhomogeneous strain.}
Such
an effect to suppress the strain distributions by an applied magnetic
field was mentioned in
\cite{acosta2013optical}, and was recently demonstrated in a vacuum Rabi
oscillation between a superconducting flux qubit and NV$^-$ centers in
\cite{matsuzaki2015improvingpra}. Our results here are consistent with
these previous results.

\subsection{Parameter estimation}
 An ensemble of NV$^-$ centers is affected by inhomogeneous magnetic
 fields, inhomogeneous strain distributions, and homogeneous broadening.
In the ODMR, the observed peaks
 contain the information of the total width that is a composite effect of
 three noise mentioned above, and so it was not straightforward to separate
 these three effects for the
 estimation about how individual noise contributes to the width.

 \textcolor{black}{
Interestingly, our model could be used to determine these three
parameters by reproducing the ODMR with and without applied magnetic fields.
Firstly, as we described before,
the sharp dip in the ODMR is very sensitive
against the change in $\Gamma _{b}$, while the dip is
relatively insensitive against the change in $\delta E_k$ and $\delta
B_k$.
 These properties are important
to determine the value of $\Gamma
_{b}$ from the analysis of the ODMR. Usually, $\Gamma
_{b}$ is much smaller than the $\delta B_k$ and $\delta
E_k$
\cite{zhu2011coherent,saito2013towards,zhudark2014,matsuzaki2015improvingpra},
and so it seems that the effect of $\Gamma
_{b}$ might be hindered by a huge influence of $\delta B_k$ and $\delta
E_k$.
However, since the dip is sensitive against the change in
$\Gamma_{b}$, we could accurately estimate the $\Gamma_{b}$ even under the effect of
$\delta B_k$ and $\delta E_k$.
Secondly, as we have shown, the ODMR
signals with applied magnetic field is robust against the
strain variations $\delta E_k$ while the peak will be broadened due to the effect of the
randomized magnetic field $\delta B_k$.
We can use these properties to estimate the $\delta B^{(k)}_z$ and
$\delta E_k$. Since the ODMR with an applied magnetic field is
insensitive against $\delta E_k$, we can estimate $\delta B^{(k)}_z$
from this experimental data. Since we have estimated $\delta B^{(k)}_z$
and $\Gamma _b$ from the prescription described above, we fix these
parameters so that we can estimate $\delta E_k$ from the ODMR with zero
applied magnetic field. Therefore, by applying these procedure, we could
estimate the parameters of the NV$^-$ centers such as inhomogeneous magnetic
 fields, inhomogeneous strain distributions, and homogeneous broadening.}

\section{Summary}
In conclusion, we have studied an ODMR with a high-density ensemble of
NV$^-$ centers.
Our model succeeds to reproduce the ODMR with and without applied
magnetic field. Also, we have shown that our model is useful to
determine
the typical parameters of the ensemble NV$^-$ centers such as
strain distributions,
 inhomogeneous magnetic fields, and homogeneous broadening width.
 Such a parameter estimation is essential for the use of NV$^-$ centers
 %toward realizing
 to realize diamond-based quantum information processing.

 Y.M thanks K. Nemoto and H. Nakano for discussion.
This work was supported 
by
JSPS
KAKENHI No. 15K17732, JSPS KAKENHI Grant No. 25220601, and the Commissioned
Research of NICT.

\textcolor{black}{\section{Appendix}}
Here, we consider the effect of inhomogeneous microwave amplitude.
If we have such an
inhomogeneity, by solving the Heisenberg equation, we obtain
\begin{eqnarray}
&&\langle \hat{b}^{\dagger }_{k,t=\infty } \hat{b}_{k,t=\infty
 }\rangle\nonumber \\
 &=& |\frac{\lambda_k(\omega -\omega ^{(k)}_d+i\Gamma _d)}{(\omega -\omega ^{(k)}_b+i\Gamma
  _b)(\omega -\omega ^{(k)}_d+i\Gamma _d) -(|J_k|^2
  +|J'_k|^2)}|^2\nonumber \\
  &&\langle \hat{d}^{\dagger }_{k,t=\infty }\hat{d}_{k,t=\infty }\rangle
   \nonumber \\
   &=&| \frac{\lambda_k(J_k-iJ'_k)}{(\omega -\omega ^{(k)}_b+i\Gamma
  _b)(\omega -\omega ^{(k)}_d+i\Gamma _d) -(|J_k|^2 +|J'_k|^2)}|^2\nonumber 
\end{eqnarray}
where the value of $\lambda $ differs depending on the position of the
NV$^-$ centers.
If we define an average probability of the NV$^-$ center in the bright (dark) state
as $P_b$ ($P_d$), we obtain
\begin{eqnarray}
 P_b&=&\sum_{k=1}^{N}\frac{|\frac{\lambda_k(\omega -\omega ^{(k)}_d+i\Gamma _d)}{(\omega -\omega ^{(k)}_b+i\Gamma
  _b)(\omega -\omega ^{(k)}_d+i\Gamma _d) -(|J_k|^2
  +|J'_k|^2)}|^2}{N}\nonumber \\
  P_d&=&\frac{\sum_{k=1}^{N}| \frac{\lambda_k(J_k-iJ'_k)}{(\omega -\omega ^{(k)}_b+i\Gamma
  _b)(\omega -\omega ^{(k)}_d+i\Gamma _d) -(|J_k|^2
  +|J'_k|^2)}|^2}{N}\nonumber \\
\end{eqnarray}
Since inhomogeneity of $\lambda $ is independent from the
inhomogeneity of $\omega _b$, $\omega _d$, $J$, and $J'$, we can
rewrite these probabilities for a large number of NV$^-$ centers as
follows
\begin{eqnarray}
 P_b&\simeq& \frac{1}{N}\sum_{j=1}^{m}|\lambda
  _j|^2\sum_{k=1}^{\text{floor}(\frac{N}{m})}
  p^{(b)}_{k}\nonumber \\
 &=&(\frac{1}{m}\sum_{j=1}^{m}|\lambda
  _j|^2)(\frac{1}{(\frac{N}{m})}\sum_{k=1}^{\text{floor}(\frac{N}{m})}
  p^{(b)}_{k})
\end{eqnarray}
\begin{eqnarray}
   P_d&\simeq &\frac{1}{N}\sum_{j=1}^{m}|\lambda
  _j|^2\sum_{k=1}^{\text{floor}(\frac{N}{m})}
  p^{(d)}_{k}\nonumber \\
 &=&(\frac{1}{m}\sum_{j=1}^{m}|\lambda
  _j|^2)(\frac{1}{(\frac{N}{m})} \sum_{k=1}^{\text{floor}(\frac{N}{m})}
  p^{(d)}_{k})
\end{eqnarray}
 where
 \begin{eqnarray}
  p^{(b)}_{k}=|\frac{(\omega -\omega ^{(k)}_d+i\Gamma _d)}{(\omega -\omega ^{(k)}_b+i\Gamma
  _b)(\omega -\omega ^{(k)}_d+i\Gamma _d) -(|J_k|^2
  +|J'_k|^2)}|^2\ \ \ \ \ \ \nonumber \\
    p^{(d)}_{k}=| \frac{(J_k-iJ'_k)}{(\omega -\omega ^{(k)}_b+i\Gamma
  _b)(\omega -\omega ^{(k)}_d+i\Gamma _d) -(|J_k|^2
  +|J'_k|^2)}|^2\nonumber 
 \end{eqnarray}
Therefore, we obtain
% $ P_b\simeq |\lambda |^2_{\text{av}}(\frac{1}{N'}\sum_{k=1}^{N'}
%   p^{(b)}_{k})$ and $  P_d\simeq 
%    |\lambda |^2_{\text{av}}(\frac{1}{N'}\sum_{k=1}^{N'}
%   p^{(d)}_{k})$
 \begin{eqnarray}
  P_b&\simeq& |\lambda |^2_{\text{av}}(\frac{1}{N'}\sum_{k=1}^{N'}
   p^{(b)}_{k})
   \\
   P_d&\simeq &
    |\lambda |^2_{\text{av}}(\frac{1}{N'}\sum_{k=1}^{N'}
   p^{(d)}_{k})
 \end{eqnarray}
where $|\lambda _{\text{av}}|^2=(\frac{1}{m}\sum_{j=1}^{m}|\lambda
  _j|^2)$ and $N'=\text{floor}(\frac{N}{m})$.
The probability of the NV$^-$ center in the ground states can be
  calculated as 
  \begin{eqnarray}
   P_e\simeq P_b+P_d
  \end{eqnarray}
and this is the same form as the probability of the homogeneous
microwave amplitude case
described in the Eq. (\ref{probabilitye}) where $N$
($\lambda ^2$) is replaced by $N'$ ($|\lambda |^2_{\text{av}}$).
So the inhomogeneous microwave amplitude does not affect the theoretical
prediction of
ODMR signals.
 
%\bibliography{5mylibrary.bib}

\begin{thebibliography}{43}
\expandafter\ifx\csname natexlab\endcsname\relax\def\natexlab#1{#1}\fi
\expandafter\ifx\csname bibnamefont\endcsname\relax
  \def\bibnamefont#1{#1}\fi
\expandafter\ifx\csname bibfnamefont\endcsname\relax
  \def\bibfnamefont#1{#1}\fi
\expandafter\ifx\csname citenamefont\endcsname\relax
  \def\citenamefont#1{#1}\fi
\expandafter\ifx\csname url\endcsname\relax
  \def\url#1{\texttt{#1}}\fi
\expandafter\ifx\csname urlprefix\endcsname\relax\def\urlprefix{URL }\fi
\providecommand{\bibinfo}[2]{#2}
\providecommand{\eprint}[2][]{\url{#2}}

\bibitem[{\citenamefont{Davies and Hamer}(1976)}]{davies1976optical}
\bibinfo{author}{\bibfnamefont{G.}~\bibnamefont{Davies}} \bibnamefont{and}
  \bibinfo{author}{\bibfnamefont{M.}~\bibnamefont{Hamer}},
  \bibinfo{journal}{Proceedings of the Royal Society of London. A. Mathematical
  and Physical Sciences} \textbf{\bibinfo{volume}{348}}, \bibinfo{pages}{285}
  (\bibinfo{year}{1976}).

\bibitem[{\citenamefont{Gruber et~al.}(1997)\citenamefont{Gruber,
  Dr{\"a}benstedt, Tietz, Fleury, Wrachtrup, and
  Von~Borczyskowski}}]{gruber1997scanning}
\bibinfo{author}{\bibfnamefont{A.}~\bibnamefont{Gruber}},
  \bibinfo{author}{\bibfnamefont{A.}~\bibnamefont{Dr{\"a}benstedt}},
  \bibinfo{author}{\bibfnamefont{C.}~\bibnamefont{Tietz}},
  \bibinfo{author}{\bibfnamefont{L.}~\bibnamefont{Fleury}},
  \bibinfo{author}{\bibfnamefont{J.}~\bibnamefont{Wrachtrup}},
  \bibnamefont{and}
  \bibinfo{author}{\bibfnamefont{C.}~\bibnamefont{Von~Borczyskowski}},
  \bibinfo{journal}{Science} \textbf{\bibinfo{volume}{276}},
  \bibinfo{pages}{2012} (\bibinfo{year}{1997}).

\bibitem[{\citenamefont{Davies}(1994)}]{Go01a}
\bibinfo{author}{\bibfnamefont{G.}~\bibnamefont{Davies}},
  \emph{\bibinfo{title}{Properties and Growth of Diamond}}
  (\bibinfo{publisher}{Inspec/Iee}, \bibinfo{year}{1994}).

\bibitem[{\citenamefont{Dutt et~al.}(2007)\citenamefont{Dutt, Childress, Jiang,
  Togan, Maze, Jelezko, Zibrov, Hemmer, and Lukin}}]{dutt2007quantum}
\bibinfo{author}{\bibfnamefont{M.~G.} \bibnamefont{Dutt}},
  \bibinfo{author}{\bibfnamefont{L.}~\bibnamefont{Childress}},
  \bibinfo{author}{\bibfnamefont{L.}~\bibnamefont{Jiang}},
  \bibinfo{author}{\bibfnamefont{E.}~\bibnamefont{Togan}},
  \bibinfo{author}{\bibfnamefont{J.}~\bibnamefont{Maze}},
  \bibinfo{author}{\bibfnamefont{F.}~\bibnamefont{Jelezko}},
  \bibinfo{author}{\bibfnamefont{A.}~\bibnamefont{Zibrov}},
  \bibinfo{author}{\bibfnamefont{P.}~\bibnamefont{Hemmer}}, \bibnamefont{and}
  \bibinfo{author}{\bibfnamefont{M.}~\bibnamefont{Lukin}},
  \bibinfo{journal}{Science} \textbf{\bibinfo{volume}{316}},
  \bibinfo{pages}{1312} (\bibinfo{year}{2007}).

\bibitem[{\citenamefont{Wrachtrup et~al.}(2001)\citenamefont{Wrachtrup, Kilin,
  and Nizovtsev}}]{wrachtrup2001quantum}
\bibinfo{author}{\bibfnamefont{J.}~\bibnamefont{Wrachtrup}},
  \bibinfo{author}{\bibfnamefont{S.~Y.} \bibnamefont{Kilin}}, \bibnamefont{and}
  \bibinfo{author}{\bibfnamefont{A.}~\bibnamefont{Nizovtsev}},
  \bibinfo{journal}{Optics and Spectroscopy} \textbf{\bibinfo{volume}{91}},
  \bibinfo{pages}{429} (\bibinfo{year}{2001}).

\bibitem[{\citenamefont{Jelezko et~al.}(2002)\citenamefont{Jelezko, Popa,
  Gruber, Tietz, Wrachtrup, Nizovtsev, and Kilin}}]{jelezko2002single}
\bibinfo{author}{\bibfnamefont{F.}~\bibnamefont{Jelezko}},
  \bibinfo{author}{\bibfnamefont{I.}~\bibnamefont{Popa}},
  \bibinfo{author}{\bibfnamefont{A.}~\bibnamefont{Gruber}},
  \bibinfo{author}{\bibfnamefont{C.}~\bibnamefont{Tietz}},
  \bibinfo{author}{\bibfnamefont{J.}~\bibnamefont{Wrachtrup}},
  \bibinfo{author}{\bibfnamefont{A.}~\bibnamefont{Nizovtsev}},
  \bibnamefont{and} \bibinfo{author}{\bibfnamefont{S.}~\bibnamefont{Kilin}},
  \bibinfo{journal}{Applied physics letters} \textbf{\bibinfo{volume}{81}},
  \bibinfo{pages}{2160} (\bibinfo{year}{2002}).

\bibitem[{\citenamefont{Neumann et~al.}(2008)\citenamefont{Neumann, Mizuochi,
  Rempp, Hemmer, Watanabe, Yamasaki, Jacques, Gaebel, Jelezko, and
  Wrachtrup}}]{NMRHWYJGJW01a}
\bibinfo{author}{\bibfnamefont{P.}~\bibnamefont{Neumann}},
  \bibinfo{author}{\bibfnamefont{N.}~\bibnamefont{Mizuochi}},
  \bibinfo{author}{\bibfnamefont{F.}~\bibnamefont{Rempp}},
  \bibinfo{author}{\bibfnamefont{P.}~\bibnamefont{Hemmer}},
  \bibinfo{author}{\bibfnamefont{H.}~\bibnamefont{Watanabe}},
  \bibinfo{author}{\bibfnamefont{S.}~\bibnamefont{Yamasaki}},
  \bibinfo{author}{\bibfnamefont{V.}~\bibnamefont{Jacques}},
  \bibinfo{author}{\bibfnamefont{T.}~\bibnamefont{Gaebel}},
  \bibinfo{author}{\bibfnamefont{F.}~\bibnamefont{Jelezko}}, \bibnamefont{and}
  \bibinfo{author}{\bibfnamefont{J.}~\bibnamefont{Wrachtrup}},
  \bibinfo{journal}{Science} \textbf{\bibinfo{volume}{320}},
  \bibinfo{pages}{1326} (\bibinfo{year}{2008}).

\bibitem[{\citenamefont{Robledo et~al.}(2011)\citenamefont{Robledo, Childress,
  Bernien, Hensen, Alkemade, and Hanson}}]{robledo2011high}
\bibinfo{author}{\bibfnamefont{L.}~\bibnamefont{Robledo}},
  \bibinfo{author}{\bibfnamefont{L.}~\bibnamefont{Childress}},
  \bibinfo{author}{\bibfnamefont{H.}~\bibnamefont{Bernien}},
  \bibinfo{author}{\bibfnamefont{B.}~\bibnamefont{Hensen}},
  \bibinfo{author}{\bibfnamefont{P.~F.} \bibnamefont{Alkemade}},
  \bibnamefont{and} \bibinfo{author}{\bibfnamefont{R.}~\bibnamefont{Hanson}},
  \bibinfo{journal}{Nature} \textbf{\bibinfo{volume}{477}},
  \bibinfo{pages}{574} (\bibinfo{year}{2011}).

\bibitem[{\citenamefont{Shimo-Oka et~al.}(2015)\citenamefont{Shimo-Oka, Kato,
  Yamasaki, Jelezko, Miwa, Suzuki, and Mizuochi}}]{shimo2015control}
\bibinfo{author}{\bibfnamefont{T.}~\bibnamefont{Shimo-Oka}},
  \bibinfo{author}{\bibfnamefont{H.}~\bibnamefont{Kato}},
  \bibinfo{author}{\bibfnamefont{S.}~\bibnamefont{Yamasaki}},
  \bibinfo{author}{\bibfnamefont{F.}~\bibnamefont{Jelezko}},
  \bibinfo{author}{\bibfnamefont{S.}~\bibnamefont{Miwa}},
  \bibinfo{author}{\bibfnamefont{Y.}~\bibnamefont{Suzuki}}, \bibnamefont{and}
  \bibinfo{author}{\bibfnamefont{N.}~\bibnamefont{Mizuochi}},
  \bibinfo{journal}{Applied Physics Letters} \textbf{\bibinfo{volume}{106}},
  \bibinfo{pages}{153103} (\bibinfo{year}{2015}).

\bibitem[{\citenamefont{Childress et~al.}(2005)\citenamefont{Childress, Taylor,
  S{\o}rensen, and Lukin}}]{CTSL01a}
\bibinfo{author}{\bibfnamefont{L.}~\bibnamefont{Childress}},
  \bibinfo{author}{\bibfnamefont{J.~M.} \bibnamefont{Taylor}},
  \bibinfo{author}{\bibfnamefont{A.~S.} \bibnamefont{S{\o}rensen}},
  \bibnamefont{and} \bibinfo{author}{\bibfnamefont{M.~D.} \bibnamefont{Lukin}},
  \bibinfo{journal}{Phys. Rev. A} \textbf{\bibinfo{volume}{72}},
  \bibinfo{pages}{052330} (\bibinfo{year}{2005}).

\bibitem[{\citenamefont{Mizuochi et~al.}(2009)\citenamefont{Mizuochi, Neumann,
  Rempp, Beck, Jacques, Siyushev, Nakamura, Twitchen, Watanabe, Yamasaki
  et~al.}}]{mizuochi2009coherence}
\bibinfo{author}{\bibfnamefont{N.}~\bibnamefont{Mizuochi}},
  \bibinfo{author}{\bibfnamefont{P.}~\bibnamefont{Neumann}},
  \bibinfo{author}{\bibfnamefont{F.}~\bibnamefont{Rempp}},
  \bibinfo{author}{\bibfnamefont{J.}~\bibnamefont{Beck}},
  \bibinfo{author}{\bibfnamefont{V.}~\bibnamefont{Jacques}},
  \bibinfo{author}{\bibfnamefont{P.}~\bibnamefont{Siyushev}},
  \bibinfo{author}{\bibfnamefont{K.}~\bibnamefont{Nakamura}},
  \bibinfo{author}{\bibfnamefont{D.}~\bibnamefont{Twitchen}},
  \bibinfo{author}{\bibfnamefont{H.}~\bibnamefont{Watanabe}},
  \bibinfo{author}{\bibfnamefont{S.}~\bibnamefont{Yamasaki}},
  \bibnamefont{et~al.}, \bibinfo{journal}{Physical review B}
  \textbf{\bibinfo{volume}{80}}, \bibinfo{pages}{041201}
  (\bibinfo{year}{2009}).

\bibitem[{\citenamefont{Balasubramanian
  et~al.}(2009)\citenamefont{Balasubramanian, Neumann, Twitchen, Markham,
  Kolesov, Mizuochi, Isoya, Achard, Beck, Tissler
  et~al.}}]{balasubramanian2009ultralong}
\bibinfo{author}{\bibfnamefont{G.}~\bibnamefont{Balasubramanian}},
  \bibinfo{author}{\bibfnamefont{P.}~\bibnamefont{Neumann}},
  \bibinfo{author}{\bibfnamefont{D.}~\bibnamefont{Twitchen}},
  \bibinfo{author}{\bibfnamefont{M.}~\bibnamefont{Markham}},
  \bibinfo{author}{\bibfnamefont{R.}~\bibnamefont{Kolesov}},
  \bibinfo{author}{\bibfnamefont{N.}~\bibnamefont{Mizuochi}},
  \bibinfo{author}{\bibfnamefont{J.}~\bibnamefont{Isoya}},
  \bibinfo{author}{\bibfnamefont{J.}~\bibnamefont{Achard}},
  \bibinfo{author}{\bibfnamefont{J.}~\bibnamefont{Beck}},
  \bibinfo{author}{\bibfnamefont{J.}~\bibnamefont{Tissler}},
  \bibnamefont{et~al.}, \bibinfo{journal}{Nature materials}
  \textbf{\bibinfo{volume}{8}}, \bibinfo{pages}{383} (\bibinfo{year}{2009}).

\bibitem[{\citenamefont{Bar-Gill et~al.}(2013)\citenamefont{Bar-Gill, Pham,
  Jarmola, Budker, and Walsworth}}]{bar2013solid}
\bibinfo{author}{\bibfnamefont{N.}~\bibnamefont{Bar-Gill}},
  \bibinfo{author}{\bibfnamefont{L.~M.} \bibnamefont{Pham}},
  \bibinfo{author}{\bibfnamefont{A.}~\bibnamefont{Jarmola}},
  \bibinfo{author}{\bibfnamefont{D.}~\bibnamefont{Budker}}, \bibnamefont{and}
  \bibinfo{author}{\bibfnamefont{R.~L.} \bibnamefont{Walsworth}},
  \bibinfo{journal}{Nature communications} \textbf{\bibinfo{volume}{4}},
  \bibinfo{pages}{1743} (\bibinfo{year}{2013}).

\bibitem[{\citenamefont{Jelezko et~al.}(2004)\citenamefont{Jelezko, Gaebel,
  Popa, Gruber, and Wrachtrup}}]{JGPGW01a}
\bibinfo{author}{\bibfnamefont{F.}~\bibnamefont{Jelezko}},
  \bibinfo{author}{\bibfnamefont{T.}~\bibnamefont{Gaebel}},
  \bibinfo{author}{\bibfnamefont{I.}~\bibnamefont{Popa}},
  \bibinfo{author}{\bibfnamefont{A.}~\bibnamefont{Gruber}}, \bibnamefont{and}
  \bibinfo{author}{\bibfnamefont{J.}~\bibnamefont{Wrachtrup}},
  \bibinfo{journal}{Phys. Rev. Lett} \textbf{\bibinfo{volume}{92}},
  \bibinfo{pages}{076401} (\bibinfo{year}{2004}).

\bibitem[{\citenamefont{Bernien et~al.}(2013)\citenamefont{Bernien, Hensen,
  Pfaff, Koolstra, Blok, Robledo, Taminiau, Markham, Twitchen, Childress
  et~al.}}]{bernien2013heralded}
\bibinfo{author}{\bibfnamefont{H.}~\bibnamefont{Bernien}},
  \bibinfo{author}{\bibfnamefont{B.}~\bibnamefont{Hensen}},
  \bibinfo{author}{\bibfnamefont{W.}~\bibnamefont{Pfaff}},
  \bibinfo{author}{\bibfnamefont{G.}~\bibnamefont{Koolstra}},
  \bibinfo{author}{\bibfnamefont{M.}~\bibnamefont{Blok}},
  \bibinfo{author}{\bibfnamefont{L.}~\bibnamefont{Robledo}},
  \bibinfo{author}{\bibfnamefont{T.}~\bibnamefont{Taminiau}},
  \bibinfo{author}{\bibfnamefont{M.}~\bibnamefont{Markham}},
  \bibinfo{author}{\bibfnamefont{D.}~\bibnamefont{Twitchen}},
  \bibinfo{author}{\bibfnamefont{L.}~\bibnamefont{Childress}},
  \bibnamefont{et~al.}, \bibinfo{journal}{Nature}
  \textbf{\bibinfo{volume}{497}}, \bibinfo{pages}{86} (\bibinfo{year}{2013}).

\bibitem[{\citenamefont{Maze et~al.}(2008)\citenamefont{Maze, Stanwix, Hodges,
  Hong, Taylor, Cappellaro, Jiang, Dutt, Togan, Zibrov
  et~al.}}]{maze2008nanoscale}
\bibinfo{author}{\bibfnamefont{J.}~\bibnamefont{Maze}},
  \bibinfo{author}{\bibfnamefont{P.}~\bibnamefont{Stanwix}},
  \bibinfo{author}{\bibfnamefont{J.}~\bibnamefont{Hodges}},
  \bibinfo{author}{\bibfnamefont{S.}~\bibnamefont{Hong}},
  \bibinfo{author}{\bibfnamefont{J.}~\bibnamefont{Taylor}},
  \bibinfo{author}{\bibfnamefont{P.}~\bibnamefont{Cappellaro}},
  \bibinfo{author}{\bibfnamefont{L.}~\bibnamefont{Jiang}},
  \bibinfo{author}{\bibfnamefont{M.}~\bibnamefont{Dutt}},
  \bibinfo{author}{\bibfnamefont{E.}~\bibnamefont{Togan}},
  \bibinfo{author}{\bibfnamefont{A.}~\bibnamefont{Zibrov}},
  \bibnamefont{et~al.}, \bibinfo{journal}{Nature}
  \textbf{\bibinfo{volume}{455}}, \bibinfo{pages}{644} (\bibinfo{year}{2008}),
  ISSN \bibinfo{issn}{0028-0836}.

\bibitem[{\citenamefont{Taylor et~al.}(2008)\citenamefont{Taylor, Cappellaro,
  Childress, Jiang, Budker, Hemmer, Yacoby, Walsworth, and
  Lukin}}]{taylor2008high}
\bibinfo{author}{\bibfnamefont{J.}~\bibnamefont{Taylor}},
  \bibinfo{author}{\bibfnamefont{P.}~\bibnamefont{Cappellaro}},
  \bibinfo{author}{\bibfnamefont{L.}~\bibnamefont{Childress}},
  \bibinfo{author}{\bibfnamefont{L.}~\bibnamefont{Jiang}},
  \bibinfo{author}{\bibfnamefont{D.}~\bibnamefont{Budker}},
  \bibinfo{author}{\bibfnamefont{P.}~\bibnamefont{Hemmer}},
  \bibinfo{author}{\bibfnamefont{A.}~\bibnamefont{Yacoby}},
  \bibinfo{author}{\bibfnamefont{R.}~\bibnamefont{Walsworth}},
  \bibnamefont{and} \bibinfo{author}{\bibfnamefont{M.}~\bibnamefont{Lukin}},
  \bibinfo{journal}{Nature Physics} \textbf{\bibinfo{volume}{4}},
  \bibinfo{pages}{810} (\bibinfo{year}{2008}).

\bibitem[{\citenamefont{Balasubramanian
  et~al.}(2008)\citenamefont{Balasubramanian, Chan, Kolesov, Al-Hmoud, Tisler,
  Shin, Kim, Wojcik, Hemmer, Krueger et~al.}}]{balasubramanian2008nanoscale}
\bibinfo{author}{\bibfnamefont{G.}~\bibnamefont{Balasubramanian}},
  \bibinfo{author}{\bibfnamefont{I.}~\bibnamefont{Chan}},
  \bibinfo{author}{\bibfnamefont{R.}~\bibnamefont{Kolesov}},
  \bibinfo{author}{\bibfnamefont{M.}~\bibnamefont{Al-Hmoud}},
  \bibinfo{author}{\bibfnamefont{J.}~\bibnamefont{Tisler}},
  \bibinfo{author}{\bibfnamefont{C.}~\bibnamefont{Shin}},
  \bibinfo{author}{\bibfnamefont{C.}~\bibnamefont{Kim}},
  \bibinfo{author}{\bibfnamefont{A.}~\bibnamefont{Wojcik}},
  \bibinfo{author}{\bibfnamefont{P.}~\bibnamefont{Hemmer}},
  \bibinfo{author}{\bibfnamefont{A.}~\bibnamefont{Krueger}},
  \bibnamefont{et~al.}, \bibinfo{journal}{Nature}
  \textbf{\bibinfo{volume}{455}}, \bibinfo{pages}{648} (\bibinfo{year}{2008}).

\bibitem[{\citenamefont{Pham et~al.}(2011)\citenamefont{Pham, Le~Sage, Stanwix,
  Yeung, Glenn, Trifonov, Cappellaro, Hemmer, Lukin, Park
  et~al.}}]{pham2011magnetic}
\bibinfo{author}{\bibfnamefont{L.~M.} \bibnamefont{Pham}},
  \bibinfo{author}{\bibfnamefont{D.}~\bibnamefont{Le~Sage}},
  \bibinfo{author}{\bibfnamefont{P.~L.} \bibnamefont{Stanwix}},
  \bibinfo{author}{\bibfnamefont{T.~K.} \bibnamefont{Yeung}},
  \bibinfo{author}{\bibfnamefont{D.}~\bibnamefont{Glenn}},
  \bibinfo{author}{\bibfnamefont{A.}~\bibnamefont{Trifonov}},
  \bibinfo{author}{\bibfnamefont{P.}~\bibnamefont{Cappellaro}},
  \bibinfo{author}{\bibfnamefont{P.}~\bibnamefont{Hemmer}},
  \bibinfo{author}{\bibfnamefont{M.~D.} \bibnamefont{Lukin}},
  \bibinfo{author}{\bibfnamefont{H.}~\bibnamefont{Park}}, \bibnamefont{et~al.},
  \bibinfo{journal}{New Journal of Physics} \textbf{\bibinfo{volume}{13}},
  \bibinfo{pages}{045021} (\bibinfo{year}{2011}).

\bibitem[{\citenamefont{Acosta et~al.}(2009)\citenamefont{Acosta, Bauch,
  Ledbetter, Santori, Fu, Barclay, Beausoleil, Linget, Roch, Treussart
  et~al.}}]{acosta2009diamonds}
\bibinfo{author}{\bibfnamefont{V.}~\bibnamefont{Acosta}},
  \bibinfo{author}{\bibfnamefont{E.}~\bibnamefont{Bauch}},
  \bibinfo{author}{\bibfnamefont{M.}~\bibnamefont{Ledbetter}},
  \bibinfo{author}{\bibfnamefont{C.}~\bibnamefont{Santori}},
  \bibinfo{author}{\bibfnamefont{K.-M.} \bibnamefont{Fu}},
  \bibinfo{author}{\bibfnamefont{P.}~\bibnamefont{Barclay}},
  \bibinfo{author}{\bibfnamefont{R.}~\bibnamefont{Beausoleil}},
  \bibinfo{author}{\bibfnamefont{H.}~\bibnamefont{Linget}},
  \bibinfo{author}{\bibfnamefont{J.}~\bibnamefont{Roch}},
  \bibinfo{author}{\bibfnamefont{F.}~\bibnamefont{Treussart}},
  \bibnamefont{et~al.}, \bibinfo{journal}{Physical Review B}
  \textbf{\bibinfo{volume}{80}}, \bibinfo{pages}{115202}
  (\bibinfo{year}{2009}).

\bibitem[{\citenamefont{Steinert et~al.}(2010)\citenamefont{Steinert, Dolde,
  Neumann, Aird, Naydenov, Balasubramanian, Jelezko, and
  Wrachtrup}}]{steinert2010high}
\bibinfo{author}{\bibfnamefont{S.}~\bibnamefont{Steinert}},
  \bibinfo{author}{\bibfnamefont{F.}~\bibnamefont{Dolde}},
  \bibinfo{author}{\bibfnamefont{P.}~\bibnamefont{Neumann}},
  \bibinfo{author}{\bibfnamefont{A.}~\bibnamefont{Aird}},
  \bibinfo{author}{\bibfnamefont{B.}~\bibnamefont{Naydenov}},
  \bibinfo{author}{\bibfnamefont{G.}~\bibnamefont{Balasubramanian}},
  \bibinfo{author}{\bibfnamefont{F.}~\bibnamefont{Jelezko}}, \bibnamefont{and}
  \bibinfo{author}{\bibfnamefont{J.}~\bibnamefont{Wrachtrup}},
  \bibinfo{journal}{Review of scientific instruments}
  \textbf{\bibinfo{volume}{81}}, \bibinfo{pages}{043705}
  (\bibinfo{year}{2010}).

\bibitem[{\citenamefont{Hardal et~al.}(2013)\citenamefont{Hardal, Xue, Shikano,
  M{\"u}stecapl{\i}o{\u{g}}lu, and Sanders}}]{hardal2013discrete}
\bibinfo{author}{\bibfnamefont{A.~{\"U}.} \bibnamefont{Hardal}},
  \bibinfo{author}{\bibfnamefont{P.}~\bibnamefont{Xue}},
  \bibinfo{author}{\bibfnamefont{Y.}~\bibnamefont{Shikano}},
  \bibinfo{author}{\bibfnamefont{{\"O}.~E.}
  \bibnamefont{M{\"u}stecapl{\i}o{\u{g}}lu}}, \bibnamefont{and}
  \bibinfo{author}{\bibfnamefont{B.~C.} \bibnamefont{Sanders}},
  \bibinfo{journal}{Physical Review A} \textbf{\bibinfo{volume}{88}},
  \bibinfo{pages}{022303} (\bibinfo{year}{2013}).

\bibitem[{\citenamefont{Yang et~al.}(2012)\citenamefont{Yang, Yin, Chen, Kou,
  Feng, and Oh}}]{yang2012quantum}
\bibinfo{author}{\bibfnamefont{W.}~\bibnamefont{Yang}},
  \bibinfo{author}{\bibfnamefont{Z.-q.} \bibnamefont{Yin}},
  \bibinfo{author}{\bibfnamefont{Z.}~\bibnamefont{Chen}},
  \bibinfo{author}{\bibfnamefont{S.-P.} \bibnamefont{Kou}},
  \bibinfo{author}{\bibfnamefont{M.}~\bibnamefont{Feng}}, \bibnamefont{and}
  \bibinfo{author}{\bibfnamefont{C.}~\bibnamefont{Oh}},
  \bibinfo{journal}{Physical Review A} \textbf{\bibinfo{volume}{86}},
  \bibinfo{pages}{012307} (\bibinfo{year}{2012}).

\bibitem[{\citenamefont{Imamo{\u{g}}lu}(2009)}]{imamouglu2009cavity}
\bibinfo{author}{\bibfnamefont{A.}~\bibnamefont{Imamo{\u{g}}lu}},
  \bibinfo{journal}{Phys. Rev. Lett.} \textbf{\bibinfo{volume}{102}},
  \bibinfo{pages}{083602} (\bibinfo{year}{2009}).

\bibitem[{\citenamefont{Wesenberg and {\it{et al}
  }}(2009)}]{wesenberg2009quantumetal}
\bibinfo{author}{\bibfnamefont{J.}~\bibnamefont{Wesenberg}} \bibnamefont{and}
  \bibinfo{author}{\bibnamefont{{\it{et al} }}}, \bibinfo{journal}{Phys. Rev.
  Lett.} \textbf{\bibinfo{volume}{103}}, \bibinfo{pages}{70502}
  (\bibinfo{year}{2009}).

\bibitem[{\citenamefont{Schuster and {\it{et al}
  }}(2010)}]{schuster2010highetal}
\bibinfo{author}{\bibfnamefont{D.}~\bibnamefont{Schuster}} \bibnamefont{and}
  \bibinfo{author}{\bibnamefont{{\it{et al} }}}, \bibinfo{journal}{Phys. Rev.
  Lett.} \textbf{\bibinfo{volume}{105}}, \bibinfo{pages}{140501}
  (\bibinfo{year}{2010}).

\bibitem[{\citenamefont{Kubo and {\it{et al} }}(2010)}]{kubo2010strongetal}
\bibinfo{author}{\bibfnamefont{Y.}~\bibnamefont{Kubo}} \bibnamefont{and}
  \bibinfo{author}{\bibnamefont{{\it{et al} }}}, \bibinfo{journal}{Phys. Rev.
  Lett.} \textbf{\bibinfo{volume}{105}}, \bibinfo{pages}{140502}
  (\bibinfo{year}{2010}).

\bibitem[{\citenamefont{Ams{\"u}ss and {\it{et al}
  }}(2011)}]{amsuss2011cavityetal}
\bibinfo{author}{\bibfnamefont{R.}~\bibnamefont{Ams{\"u}ss}} \bibnamefont{and}
  \bibinfo{author}{\bibnamefont{{\it{et al} }}}, \bibinfo{journal}{Phys. Rev.
  Lett.} \textbf{\bibinfo{volume}{107}}, \bibinfo{pages}{060502}
  (\bibinfo{year}{2011}).

\bibitem[{\citenamefont{Marcos and {\it{et al}
  }}(2010)}]{marcos2010couplingetal}
\bibinfo{author}{\bibfnamefont{D.}~\bibnamefont{Marcos}} \bibnamefont{and}
  \bibinfo{author}{\bibnamefont{{\it{et al} }}}, \bibinfo{journal}{Phys. Rev.
  Lett.} \textbf{\bibinfo{volume}{105}}, \bibinfo{pages}{210501}
  (\bibinfo{year}{2010}).

\bibitem[{\citenamefont{Julsgaard and {\it{et al}
  }}(2013)}]{julsgaard2013quantumetal}
\bibinfo{author}{\bibfnamefont{B.}~\bibnamefont{Julsgaard}} \bibnamefont{and}
  \bibinfo{author}{\bibnamefont{{\it{et al} }}}, \bibinfo{journal}{Phys. Rev.
  Lett.} \textbf{\bibinfo{volume}{110}}, \bibinfo{pages}{250503}
  (\bibinfo{year}{2013}).

\bibitem[{\citenamefont{Diniz and {\it{et al} }}(2011)}]{diniz2011stronglyetal}
\bibinfo{author}{\bibfnamefont{I.}~\bibnamefont{Diniz}} \bibnamefont{and}
  \bibinfo{author}{\bibnamefont{{\it{et al} }}}, \bibinfo{journal}{Phys. Rev.
  A} \textbf{\bibinfo{volume}{84}}, \bibinfo{pages}{063810}
  (\bibinfo{year}{2011}).

\bibitem[{\citenamefont{Putz et~al.}(2014)\citenamefont{Putz, Krimer,
  Ams{\"u}ss, Valookaran, N{\"o}bauer, Schmiedmayer, Rotter, and
  Majer}}]{putz2014protecting}
\bibinfo{author}{\bibfnamefont{S.}~\bibnamefont{Putz}},
  \bibinfo{author}{\bibfnamefont{D.~O.} \bibnamefont{Krimer}},
  \bibinfo{author}{\bibfnamefont{R.}~\bibnamefont{Ams{\"u}ss}},
  \bibinfo{author}{\bibfnamefont{A.}~\bibnamefont{Valookaran}},
  \bibinfo{author}{\bibfnamefont{T.}~\bibnamefont{N{\"o}bauer}},
  \bibinfo{author}{\bibfnamefont{J.}~\bibnamefont{Schmiedmayer}},
  \bibinfo{author}{\bibfnamefont{S.}~\bibnamefont{Rotter}}, \bibnamefont{and}
  \bibinfo{author}{\bibfnamefont{J.}~\bibnamefont{Majer}},
  \bibinfo{journal}{Nature Physics} \textbf{\bibinfo{volume}{10}},
  \bibinfo{pages}{720} (\bibinfo{year}{2014}).

\bibitem[{\citenamefont{Zhu et~al.}(2011)\citenamefont{Zhu, Saito, Kemp,
  Kakuyanagi, Karimoto, Nakano, Munro, Tokura, Everitt, Nemoto
  et~al.}}]{zhu2011coherent}
\bibinfo{author}{\bibfnamefont{X.}~\bibnamefont{Zhu}},
  \bibinfo{author}{\bibfnamefont{S.}~\bibnamefont{Saito}},
  \bibinfo{author}{\bibfnamefont{A.}~\bibnamefont{Kemp}},
  \bibinfo{author}{\bibfnamefont{K.}~\bibnamefont{Kakuyanagi}},
  \bibinfo{author}{\bibfnamefont{S.}~\bibnamefont{Karimoto}},
  \bibinfo{author}{\bibfnamefont{H.}~\bibnamefont{Nakano}},
  \bibinfo{author}{\bibfnamefont{W.}~\bibnamefont{Munro}},
  \bibinfo{author}{\bibfnamefont{Y.}~\bibnamefont{Tokura}},
  \bibinfo{author}{\bibfnamefont{M.}~\bibnamefont{Everitt}},
  \bibinfo{author}{\bibfnamefont{K.}~\bibnamefont{Nemoto}},
  \bibnamefont{et~al.}, \bibinfo{journal}{Nature}
  \textbf{\bibinfo{volume}{478}}, \bibinfo{pages}{221} (\bibinfo{year}{2011}).

\bibitem[{\citenamefont{Zhu et~al.}(2014)\citenamefont{Zhu, Matsuzaki,
  Ams{\"u}ss, Kakuyanagi, Shimo-Oka, Mizuochi, Nemoto, Semba, Munro, and
  Saito}}]{zhudark2014}
\bibinfo{author}{\bibfnamefont{X.}~\bibnamefont{Zhu}},
  \bibinfo{author}{\bibfnamefont{Y.}~\bibnamefont{Matsuzaki}},
  \bibinfo{author}{\bibfnamefont{R.}~\bibnamefont{Ams{\"u}ss}},
  \bibinfo{author}{\bibfnamefont{K.}~\bibnamefont{Kakuyanagi}},
  \bibinfo{author}{\bibfnamefont{T.}~\bibnamefont{Shimo-Oka}},
  \bibinfo{author}{\bibfnamefont{N.}~\bibnamefont{Mizuochi}},
  \bibinfo{author}{\bibfnamefont{K.}~\bibnamefont{Nemoto}},
  \bibinfo{author}{\bibfnamefont{K.}~\bibnamefont{Semba}},
  \bibinfo{author}{\bibfnamefont{W.~J.} \bibnamefont{Munro}}, \bibnamefont{and}
  \bibinfo{author}{\bibfnamefont{S.}~\bibnamefont{Saito}},
  \bibinfo{journal}{Nature communications} \textbf{\bibinfo{volume}{5}}
  (\bibinfo{year}{2014}).

\bibitem[{\citenamefont{Kubo and {\it{et al} }}(2011)}]{kubo2011hybridetal}
\bibinfo{author}{\bibfnamefont{Y.}~\bibnamefont{Kubo}} \bibnamefont{and}
  \bibinfo{author}{\bibnamefont{{\it{et al} }}}, \bibinfo{journal}{Phys. Rev.
  Lett.} \textbf{\bibinfo{volume}{107}}, \bibinfo{pages}{220501}
  (\bibinfo{year}{2011}).

\bibitem[{\citenamefont{Saito et~al.}(2013)\citenamefont{Saito, Zhu,
  Ams{\"u}ss, Matsuzaki, Kakuyanagi, Shimo-Oka, Mizuochi, Nemoto, Munro, and
  Semba}}]{saito2013towards}
\bibinfo{author}{\bibfnamefont{S.}~\bibnamefont{Saito}},
  \bibinfo{author}{\bibfnamefont{X.}~\bibnamefont{Zhu}},
  \bibinfo{author}{\bibfnamefont{R.}~\bibnamefont{Ams{\"u}ss}},
  \bibinfo{author}{\bibfnamefont{Y.}~\bibnamefont{Matsuzaki}},
  \bibinfo{author}{\bibfnamefont{K.}~\bibnamefont{Kakuyanagi}},
  \bibinfo{author}{\bibfnamefont{T.}~\bibnamefont{Shimo-Oka}},
  \bibinfo{author}{\bibfnamefont{N.}~\bibnamefont{Mizuochi}},
  \bibinfo{author}{\bibfnamefont{K.}~\bibnamefont{Nemoto}},
  \bibinfo{author}{\bibfnamefont{W.~J.} \bibnamefont{Munro}}, \bibnamefont{and}
  \bibinfo{author}{\bibfnamefont{K.}~\bibnamefont{Semba}},
  \bibinfo{journal}{Phys. Rev. Lett.} \textbf{\bibinfo{volume}{111}},
  \bibinfo{pages}{107008} (\bibinfo{year}{2013}).

\bibitem[{\citenamefont{Alegre et~al.}(2007)\citenamefont{Alegre, Santori,
  Medeiros-Ribeiro, and Beausoleil}}]{alegre2007polarization}
\bibinfo{author}{\bibfnamefont{T.~P.~M.} \bibnamefont{Alegre}},
  \bibinfo{author}{\bibfnamefont{C.}~\bibnamefont{Santori}},
  \bibinfo{author}{\bibfnamefont{G.}~\bibnamefont{Medeiros-Ribeiro}},
  \bibnamefont{and} \bibinfo{author}{\bibfnamefont{R.~G.}
  \bibnamefont{Beausoleil}}, \bibinfo{journal}{Physical Review B}
  \textbf{\bibinfo{volume}{76}}, \bibinfo{pages}{165205}
  (\bibinfo{year}{2007}).

\bibitem[{\citenamefont{Fang et~al.}(2013)\citenamefont{Fang, Acosta, Santori,
  Huang, Itoh, Watanabe, Shikata, and Beausoleil}}]{fang2013high}
\bibinfo{author}{\bibfnamefont{K.}~\bibnamefont{Fang}},
  \bibinfo{author}{\bibfnamefont{V.~M.} \bibnamefont{Acosta}},
  \bibinfo{author}{\bibfnamefont{C.}~\bibnamefont{Santori}},
  \bibinfo{author}{\bibfnamefont{Z.}~\bibnamefont{Huang}},
  \bibinfo{author}{\bibfnamefont{K.~M.} \bibnamefont{Itoh}},
  \bibinfo{author}{\bibfnamefont{H.}~\bibnamefont{Watanabe}},
  \bibinfo{author}{\bibfnamefont{S.}~\bibnamefont{Shikata}}, \bibnamefont{and}
  \bibinfo{author}{\bibfnamefont{R.~G.} \bibnamefont{Beausoleil}},
  \bibinfo{journal}{Phys. Rev. Lett.} \textbf{\bibinfo{volume}{110}},
  \bibinfo{pages}{130802} (\bibinfo{year}{2013}).

\bibitem[{\citenamefont{Dolde et~al.}(2011)\citenamefont{Dolde, Fedder,
  Doherty, N{\"o}bauer, Rempp, Balasubramanian, Wolf, Reinhard, Hollenberg,
  Jelezko et~al.}}]{dolde2011electric}
\bibinfo{author}{\bibfnamefont{F.}~\bibnamefont{Dolde}},
  \bibinfo{author}{\bibfnamefont{H.}~\bibnamefont{Fedder}},
  \bibinfo{author}{\bibfnamefont{M.}~\bibnamefont{Doherty}},
  \bibinfo{author}{\bibfnamefont{T.}~\bibnamefont{N{\"o}bauer}},
  \bibinfo{author}{\bibfnamefont{F.}~\bibnamefont{Rempp}},
  \bibinfo{author}{\bibfnamefont{G.}~\bibnamefont{Balasubramanian}},
  \bibinfo{author}{\bibfnamefont{T.}~\bibnamefont{Wolf}},
  \bibinfo{author}{\bibfnamefont{F.}~\bibnamefont{Reinhard}},
  \bibinfo{author}{\bibfnamefont{L.}~\bibnamefont{Hollenberg}},
  \bibinfo{author}{\bibfnamefont{F.}~\bibnamefont{Jelezko}},
  \bibnamefont{et~al.}, \bibinfo{journal}{Nature Physics}
  \textbf{\bibinfo{volume}{7}}, \bibinfo{pages}{459} (\bibinfo{year}{2011}).

\bibitem[{\citenamefont{Simanovskaia et~al.}(2013)\citenamefont{Simanovskaia,
  Jensen, Jarmola, Aulenbacher, Manson, and
  Budker}}]{simanovskaia2013sidebands}
\bibinfo{author}{\bibfnamefont{M.}~\bibnamefont{Simanovskaia}},
  \bibinfo{author}{\bibfnamefont{K.}~\bibnamefont{Jensen}},
  \bibinfo{author}{\bibfnamefont{A.}~\bibnamefont{Jarmola}},
  \bibinfo{author}{\bibfnamefont{K.}~\bibnamefont{Aulenbacher}},
  \bibinfo{author}{\bibfnamefont{N.}~\bibnamefont{Manson}}, \bibnamefont{and}
  \bibinfo{author}{\bibfnamefont{D.}~\bibnamefont{Budker}},
  \bibinfo{journal}{Physical Review B} \textbf{\bibinfo{volume}{87}},
  \bibinfo{pages}{224106} (\bibinfo{year}{2013}).

\bibitem[{\citenamefont{Kubo and {\it{et al} }}(2012)}]{kubo2012electronetal}
\bibinfo{author}{\bibfnamefont{Y.}~\bibnamefont{Kubo}} \bibnamefont{and}
  \bibinfo{author}{\bibnamefont{{\it{et al} }}}, \bibinfo{journal}{Phys. Rev.
  B} \textbf{\bibinfo{volume}{86}}, \bibinfo{pages}{064514}
  (\bibinfo{year}{2012}).

\bibitem[{\citenamefont{Acosta et~al.}(2013)\citenamefont{Acosta, Budker,
  Hemmer, Maze, and Walsworth}}]{acosta2013optical}
\bibinfo{author}{\bibfnamefont{V.}~\bibnamefont{Acosta}},
  \bibinfo{author}{\bibfnamefont{D.}~\bibnamefont{Budker}},
  \bibinfo{author}{\bibfnamefont{P.}~\bibnamefont{Hemmer}},
  \bibinfo{author}{\bibfnamefont{J.}~\bibnamefont{Maze}}, \bibnamefont{and}
  \bibinfo{author}{\bibfnamefont{R.}~\bibnamefont{Walsworth}},
  \bibinfo{journal}{Optical magnetometry with nitrogen-vacancy centers in
  diamond (Cambridge University Press, Cambridge, 2013).}
  (\bibinfo{year}{2013}).

\bibitem[{\citenamefont{Matsuzaki et~al.}(2015)\citenamefont{Matsuzaki, Zhu,
  Kakuyanagi, Toida, Shimooka, Mizuochi, Nemoto, Semba, Munro, Yamaguchi
  et~al.}}]{matsuzaki2015improvingpra}
\bibinfo{author}{\bibfnamefont{Y.}~\bibnamefont{Matsuzaki}},
  \bibinfo{author}{\bibfnamefont{X.}~\bibnamefont{Zhu}},
  \bibinfo{author}{\bibfnamefont{K.}~\bibnamefont{Kakuyanagi}},
  \bibinfo{author}{\bibfnamefont{H.}~\bibnamefont{Toida}},
  \bibinfo{author}{\bibfnamefont{T.}~\bibnamefont{Shimooka}},
  \bibinfo{author}{\bibfnamefont{N.}~\bibnamefont{Mizuochi}},
  \bibinfo{author}{\bibfnamefont{K.}~\bibnamefont{Nemoto}},
  \bibinfo{author}{\bibfnamefont{K.}~\bibnamefont{Semba}},
  \bibinfo{author}{\bibfnamefont{W.}~\bibnamefont{Munro}},
  \bibinfo{author}{\bibfnamefont{H.}~\bibnamefont{Yamaguchi}},
  \bibnamefont{et~al.}, \bibinfo{journal}{Physical Review A}
  \textbf{\bibinfo{volume}{91}}, \bibinfo{pages}{042329}
  (\bibinfo{year}{2015}).

\end{thebibliography}

\end{document}